\documentclass[useAMS,usenatbib]{mn2e}
\usepackage{psfig}

\def\ltsima{$\; \buildrel < \over \sim \;$}
\def\lsim{\lower.5ex\hbox{\ltsima}}
\def\gtsima{$\; \buildrel > \over \sim \;$}
\def\gsim{\lower.5ex\hbox{\gtsima}}
\def\be{\begin{equation}}
\def\ee{\end{equation}}
\def\no{\noindent}
\def\Mesz{M\'esz\'aros}

\def\Grel{\Gamma_{\rm rel}}
\def\brel{\beta_{\rm rel}}
\begin{document}

\title{Neutron-loaded outflows in gamma-ray bursts}

\author[Rossi et al.]{Elena M. Rossi $^{1,2}$
  Andrei M. Beloborodov$^{3,4}$ 
\& Martin J. Rees$^1$\\
$^1$Institute of Astronomy, University of Cambridge, Madingley Road,
Cambridge CB3 0HA, England \\ 
$^2$Max Planck Institute for Astrophysics, Garching, 
Karl-Schwarzschild-Str. 1, D-85741 Garching, Germany \\
$^3$Physics Department and Columbia Astrophysics Laboratory, 
Columbia University, 538 W 120th Street, New York, NY 10027, USA \\
$^4$Astro-Space Center of Lebedev Physical Institute,
Profsojuznaja 84/32, Moscow 117810, Russia \\
\tt e-mail: emr@jilau1.colorado.edu, mjr@ast.cam.ac.uk amb@phys.columbia.edu}

\maketitle

\begin{abstract}
Relativistic neutron-loaded outflows in gamma-ray bursts are studied 
at their early stages, before deceleration by a surrounding medium. 
The outflow has four components: radiation, electrons, protons and neutrons.
The components interact with each other and exchange energy as the outflow 
expands. The presence of neutrons significantly changes the outflow evolution. 
Before neutrons decouple from protons, friction between the two
components increases their temperatures by many orders of magnitude. 
After the decoupling, the gradual neutron decay inside the outflow has 
a drag effect on the protons and reduces their final Lorentz factor.
\end{abstract}

\medskip

\section{Introduction}
\label{sec:introc4}

A neutron component in $\gamma$-ray bursts (GRBs) was proposed by 
Derishev, Kocharovsky \& Kocharovsky (1999a,b), and detailed calculations of 
nuclear composition show that free neutrons are inevitably present 
among ejected baryons (Beloborodov 2003b; hereafter B03b). 
Any plausible central engine of GRBs is dense and at least 
mildly degenerate, which leads to its neutronization. 
During the explosion, the expanding neutron-rich material may undergo
nucleosynthesis: neutrons tend to recombine with protons to 
$\alpha$-particles (Lemoine 2002; Pruet, Guiles \& Fuller 2002; B03b).
This recombination may be successful if the outflow is collimated.
However, even in an extreme case of a very efficient recombination, 
a significant neutron component is left over in neutron-rich outflows because 
the formation of $\alpha$-particles consumes equal numbers of neutrons and 
protons. The abundance of leftover neutrons may vary from 
10\% to more than 90\% depending on the precise parameters of the burst.

The presence of neutrons changes the theoretical picture of GRB explosion. 
Firstly, they may develop a somewhat smaller Lorentz factor than protons. 
When such a decoupling takes place, the last $n$-$p$ collisions lead to 
emission of observable multi-GeV neutrinos (Derishev et al. 1999a, hereafter 
DKK99a; Bahcall \& \Mesz~2000; \Mesz~ \& Rees 2000a).
Secondly, neutrons decay with time. The decay impacts the external blast wave 
at radii $r\sim 10^{16}-10^{17}$~cm because even an exponentially small 
number of survived neutrons carry an energy much larger than the rest-energy 
of external medium (Beloborodov 2003a).

In the present paper, we study the dynamics of neutron-loaded outflows at 
early stages of their expansion, $r<10^{16}$~cm, before they are decelerated 
by an external medium.  
The neutrons decay gradually at all radii $r\lsim 10^{17}$~cm, and at small 
$r$ the decay occurs {\it inside} the GRB outflow. The decay turns out 
important at $r$ as small as $10^{12}$~cm.

Interesting effects also take place at small $r<10^{12}$~cm.
Neutrons are initially accelerated together with protons because they are 
collisionally coupled, and the last $n$-$p$ collisions before decoupling
cause a significant heating.

We assume in this paper a simple hydrodynamic picture of expansion driven by 
thermal pressure 
and study the basic dynamics of a uniform neutron-loaded outflow.
We do not consider internal shocks or possible dynamical effects of 
magnetic fields.
The paper is organized as follows. In section~\ref{sec:pfire} we briefly 
review neutron-free outflows, which have been studied previously
in detail (see Piran 2004 for a review). In section~\ref{sec:npfire}
we derive equations describing neutron-loaded outflows and calculate 
example numerical models.
Results are discussed in section~\ref{sec:discc4}.


\section{Neutron-free outflow}
\label{sec:pfire}

We model the GRB outflow as a steady wind with duration $t_{GRB}$,
luminosity $L$, and baryon mass outflow rate $\dot{M}$. We assume spherical
symmetry. This is a good approximation also for jets with constant
opening angle greater than $1/\Gamma$ -- such a jet behaves as a part 
of a spherically symmetric outflow. 
The main parameter of the problem is
\be
 \eta=\frac{L}{\dot{M}\,c^{2}}\sim 10^2-10^3.
 \label{eq:eta}
\ee

We consider three components in this section: radiation, electrons and
protons. At small radii $r\lsim 30R_0$, the outflow is dominated by 
$e^\pm$ 
pairs, all components maintain a common temperature
and cool adiabatically. An interesting evolution begins 
at $r>50 R_0$ where $e^\pm$ can be neglected.

\subsection{Opaque stage}

\subsubsection{Outflow acceleration}

As long as the outflow is optically thick, electrons, protons and
radiation behave as a single relativistic fluid. The fluid has 
four-velocity $U^{\alpha}=(\Gamma_p c, \Gamma_p {{\mathbf\beta}_p} c,0,0)$
in spherical coordinates $(t,r,\theta,\phi)$. 
The trapped radiation is isotropic in the fluid frame 
and described by the blackbody law with a temperature $T_r$. 

The total stress energy tensor of the fluid is then given by (e.g. Misner,
Thorne \& Wheeler 1973) 
\be
{\mathcal T}^{\alpha\,\beta}=\left(\frac{4}{3} a\,T_r^{4}+\rho_p\,c^2\right) 
  \frac{U^{\alpha}U^{\beta}}{c^2}+\frac{1}{3}\,aT_r^{4} g^{\alpha\,\beta},
\label{eq:setp}
\ee
where $a=7.56\times 10^{-15}$~erg~cm$^{-3}$ $K^{-4}$ is the radiation 
constant and $g^{\alpha\beta}={\rm diag}(-1,1,r^2,r^2\sin^2\theta)$ is 
Minkowski metric in coordinates $(t,r,\theta,\phi)$. The plasma 
contribution to the energy density is taken equal to $\rho_p\,c^{2}$ where 
$\rho_p$ is the proper mass density of protons; small contributions from 
$e$ and $p$ thermal motions and the electron rest mass are neglected. 
The four-vector of baryon mass flux is given by,
\be
F^\alpha=\rho_p\,U^\alpha.
\label{eq:mfluxp}
\ee

The outflow dynamics is governed by the conservation laws,
\be
\frac{\partial {\mathcal T}^{tt}}{\partial t}
  =-\frac{1}{r^2}\frac{\partial (r^{2}\, {\mathcal T}^{tr})}{\partial r},
\label{eq:enect}
\ee
\be
\frac{\partial F^{t}}{\partial t}
  =-\frac{1}{r^{2}}\frac{\partial (r^2\,F^r)}{\partial r}.
\label{eq:mcnt}
\ee
As long as $r<t_{GRB}\,c\,\Gamma_p^{2}$ (which we assume to be satisfied 
thereafter) the outflow may be described as a steady-state wind, so that 
$\partial/\partial t=0$.\footnote{At radii $r<t_{GRB}\,c\,\Gamma_p^{2}$
the leading and trailing parts of the outflow are causially disconnected.
Therefore the outflow behaves as part of a steady wind despite the fact 
that geometrically it is a thin shell already at $r>t_{GRB}\,c$.} 
Then the conservation laws give
\be
L= 4\pi \,r^{2} c\beta_p \Gamma_p^{2} \left(\frac{4}{3} a\,T_r^{4}
   +\rho_p\,c^2\right)=const,
\label{eq:ce}
\ee
\be
\dot{M}=4\pi \,r^{2} c\beta_p \Gamma_p \rho_p=const.
\label{eq:cm}
\ee
The ratio of these two constants is the parameter $\eta$ (eq.~\ref{eq:eta}). 

The description of fluid dynamics will be complete once we specify
the evolution of radiation temperature $T_r$ with radius.
The photon-to-baryon ratio in GRB outflows is $\sim 10^5$ (see e.g. eq.~68 
in B03b) and their internal energy is strongly dominated by radiation.
Energy exchange between radiation and plasma practically does not affect 
$T_r$ and it follows the adiabatic law,
\be
T_r=T_0\,\left(\frac{n}{n_0}\right)^{1/3},
\label{eq:adlaw}
\ee
\be
 n=n_0\,\left(\frac{r}{R_0}\right)^{-2}\frac{1}{\Gamma_p\beta_p},
\label{eq:n}
\ee
where $n=\rho_p/m_p$ is the proton number density; $T_0$ and $n_0$ are
constants defined at the base of the outflow $r=R_0\sim 10^6-10^7$~cm,
\be
T_0\equiv\left(\frac{3\,L}{16\pi\Gamma_0^2\beta_0
\,R_0^{2}\,c\,a}\right)^{1/4}
  \approx 10^{10}\,\frac{L_{52}^{1/4}}{R_{07}^{1/2}}\, {\rm K},
\label{eq:temp0}
\ee
\be
n_0\equiv\frac{L}{4\pi\,\Gamma_0\beta_0R_0^2m_p\,c^3\,\eta}
   \approx 6\times 10^{26} \frac{L_{52}}{R^2_{07}}
    \left(\frac{\eta}{300}\right)^{-1}
    {\rm cm}^{-3},
\label{eq:n0}
\ee
where $\Gamma_0=\Gamma_p(R_0)$ and we assume 
\be
 \Gamma_0\beta_0=1.
\ee

Equations~(\ref{eq:ce}), (\ref{eq:cm}) and (\ref{eq:adlaw}) give
a closed description of the outflow dynamics at the opaque stage of
its expansion. Combining the equations we find
\be
  \Gamma_p(1+x)=\eta,  
\label{eq:Gx}
\ee
where
\be
   x\equiv\frac{4}{3}\frac{aT_r^4}{nm_pc^2}
         =x_0\left(\frac{n}{n_0}\right)^{1/3},
    \qquad x_0=\frac{\eta}{\Gamma_0}-1,
\label{eq:x}
\ee
and obtain the algebraic equation for $\Gamma_p(r)$,
\be 
  \left(\frac{\eta}{\Gamma_p}-1\right)\left(\Gamma_p^2-1\right)^{1/6}
    =x_0\left(\frac{r}{R_0}\right)^{-2/3},
\label{eq:gr}
\ee
(see also DKK99a).

\subsubsection{Thermal balance}

The evolution of electron temperature $T_e$ and proton temperature 
$T_p$ in general depends on their energy exchange with each other and 
radiation. It turns out that all components of the neutron-free outflow 
maintain the common temperature $T_e\approx T_p\approx T_r$ during the 
opaque stage. We shall verify this with an accurate thermodynamic 
calculation.

The electron and proton components obey the first law of thermodynamics,
\begin{equation}
 dU_i=dQ_i+(U_i+P_i)\frac{dn}{n}, \qquad i=e,p,
\label{eq:Ilaw}
\end{equation}
where $U_i$ is internal energy density of component $i$, 
$dQ_i$ is heat received by component $i$ per unit volume and $P_i$ is 
its pressure; all the quantities are measured in the fluid frame. 
$U_i$ and $P_i$ are related to temperature $T_i$,
\be
  U_i=\frac{nkT_i}{\hat{\gamma}_i-1}, \qquad P_i=nkT_i, \qquad i=e,p,
\label{eq:UP}
\ee
where $\hat{\gamma}_i$ is the adiabatic index of component $i$;
it equals $5/3$ for nonrelativistic electrons and protons.
Then from equation~(\ref{eq:Ilaw}) we obtain the equation for $T_i$, 
\begin{equation}
\frac{3}{2}\frac{dT_i}{dr}=\frac{T_i}{n}\,\frac{dn}{dr}
    +\frac{1}{kn}\,\frac{dQ_i}{dt^\prime}\,\frac{1}{c\,\Gamma_p\,\beta_p},
  \quad i=e,p,
\label{eq:dte}
\end{equation}
where $dt^\prime=dt/\Gamma_p$ is time measured in the fluid frame.
Electrons exchange energy with photons via Compton scattering and
with protons via Coulomb collisions,
\begin{equation}
\frac{dQ_e}{dt'}=\frac{3}{2}nk\left(\frac{T_C-T_e}{\tau_c}
                 +\frac{T_p-T_e}{\tau_{ep}}\right),
\label{eq:dqe}
\end{equation}
where
\be
\tau_{ep}= \sqrt{\frac{\pi}{2}}\,\frac{m_p}{m_e}\frac{1} {\sigma_T\,c\,\ln\Lambda
}\frac{1} {n_e}\left(\frac{k\,T_e} {m_ec^2}+\frac{k\,T_p} {m_pc^2}\right)^{3/2},
\label{eq:tauep}
\ee
is the Coulomb timescale (with Coulomb logarithm $\ln\Lambda\simeq 15$; see
Stepney 1983), 
\be
 \tau_C=\frac{3\,m_e\,c}{8\,U_{rad}\sigma_T},
\label{eq:tauc}
\ee
is the Compton timescale, 
$$
   U_{rad}=a\,T^4_r,
$$ 
is the radiation energy density, and $T_C$ is the Compton temperature
of radiation. For blackbody radiation $T_C=T_r$.

The thermal balance for protons reads
\begin{equation}
\frac{dQ_p}{dt^\prime}=-\frac{3}{2}nk\frac{(T_p-T_e)}{\tau_{ep}}.
\label{eq:dqp}
\end{equation}
We have neglected here the energy exchange between protons and radiation;
it is $\sim(m_e/m_p)^{3}$ times the electron-radiation energy exchange.

\subsection{Transparent stage}
\label{sec:thinp}

Electron scattering dominates opacity of the outflow, and its 
optical depth is 
\begin{equation}
 \tau_T=\frac{n_e\,\sigma_T\,r}{\Gamma_p}
       =\frac{L\,\sigma_T}{4\pi \,r m_p c^3\,\Gamma_p^2\,\eta},
\label{eq:tauw}
\end{equation}
where $n_e=n$ is the electron number density. At radii $r>R_\tau$ where 
$\tau_T<1$, the outflow is transparent and equation~(\ref{eq:adlaw}) is 
no longer valid. The photon luminosity at this point is 
(see eq.~\ref{eq:ce}),
\be
  L_\gamma=\frac{16\pi}{3} R_\tau^2 c\beta_p\Gamma_p^2aT_r^4,
\label{eq:Lg}
\ee 
where all quantities are taken at $r=R_\tau$. Approximately this 
luminosity escapes to a distant observer as a blackbody radiation
with observed temperature $\Gamma_pT_r(R_\tau)$.

\subsubsection{Outflow acceleration}

After the transparency radius, the outflow may still be accelerated by 
radiation.

A freely propagating photon at an angle $\theta$ with respect to 
radius satisfies the relation $r\sin\theta=const$ and becomes more 
beamed at larger $r$. A typical photon is emitted at angle $\pi/2$ 
in the plasma frame and has initial beaming angle 
$\theta_{rad}\approx 1/\Gamma_p$ at $r=R_\tau$. At larger 
radii the beaming angle decreases as 
\be
  \theta_{rad}\approx\frac{R_\tau}{r}\frac{1}{\Gamma_p(R_\tau)}.
\ee
One can define a frame where the freely streaming radiation remains 
approximately isotropic.
The velocity of this frame is $\beta_{rad}=\cos\theta_{rad}$ and its 
Lorentz factor is 
\be
  \Gamma_{rad}=\frac{1}{\sin\theta_{rad}}
        \approx\frac{r}{R_\tau}\Gamma_p(R_\tau).
\label{eq:Grad}
\ee
The streaming radiation will tend to accelerate the plasma outflow 
since $\Gamma_{rad}>\Gamma_p$. The radiation flux
$F_\gamma=L_\gamma/4\pi r^2$ exerts a force on an
electron moving with velocity $\beta_p$ (see Beloborodov 2002, eq. A6),
\be
  \frac{dp}{dt}=\frac{\sigma_T F_\gamma}{c}
  \left(\frac{1-\beta_p}{1+\beta_p}\right)
  \left(1-\frac{\Gamma_p^4}{\Gamma_{rad}^4}\right)
\ee
and the plasma Lorentz factor grows,
\be
 \frac{d\Gamma_p}{dr}=\frac{\sigma_TL_\gamma}{16\pi r^2m_pc^3\Gamma_p^2}
                      \left(1-\frac{\Gamma_p^4}{\Gamma_{rad}^4}\right).
\label{eq:radacc}
\ee
An easy estimate of the acceleration effect may be obtained neglecting the 
factor $\frac{\Gamma_p^4}{\Gamma_{rad}^4}$ in parenthesis. Then equation~(\ref{eq:radacc}) gives
\be
  \Gamma_p^3(r)=\Gamma_p^3(R_\tau)
   +\frac{3\sigma_TL_\gamma}{16\pi m_pc^3}
   \left(\frac{1}{R_\tau}-\frac{1}{r}\right).
\label{eq:g>rt}
\ee
The acceleration at the transparent stage is significant if at $r\gg R_\tau$ 
the second term on right-hand side exceeds the first term.
This condition is equivalent to $aT_r^4>nm_pc^2$ at $r=R_\tau$,
which requires most of the outflow energy be in 
radiation at the moment of transparency. The same condition may be 
expressed in terms of $\eta$ (e.g. M\'esz\'aros \& Rees 2000b),
\be
 \eta>\eta_{rad}\approx
   \left(\frac{L \sigma_T\,} {4\pi\,R_0 m_p c^3}\right)^{1/4} 
            \approx 10^{3} \,L^{1/4}_{52}\,R_{0,7}^{-1/4}.
 \label{eq:etamax}
\ee

\subsubsection{Thermal balance}

The thermal balance of the outflow at the transparent stage is still
given by equations~(\ref{eq:dte}), (\ref{eq:dqe}) and (\ref{eq:dqp}). 
The Coulomb timescale is given by the same equation~(\ref{eq:tauep})
and the Compton timescale is given by equation~(\ref{eq:tauc}) with
\be
 U_{rad}(r)\approx\frac{L_\gamma}{4\pi r^2 c\Gamma_p^2}
\ee
being the radiation density measured in the plasma frame.

We note that the outflow is subject to significant Compton cooling
even at $r>R_\tau$ because the scattering rate {\it per electron} is 
still very high at $R_\tau$ (it is $n_\gamma/n\sim 10^5$ higher than 
the scattering rate per photon).
The Compton temperature $T_C$ appearing in equation~(\ref{eq:dqe}) 
represents the effective temperature of radiation observed from the plasma 
frame and is proportional to the average photon energy in this frame.
$T_C$ may be evaluated as follows.

The typical photon at $r>R_\tau$ has angle $\theta(r)=\theta_{rad}(r)$
and constant energy $\epsilon(r)=\epsilon^\prime(R_\tau)\Gamma_p(R_\tau)$
in the lab frame. Its energy in the plasma frame is
\begin{eqnarray}
\nonumber
 \epsilon^\prime(r)=\epsilon\Gamma_p(1-\beta_p\cos\theta_{rad}) 
                =\epsilon\Gamma_p(1-\beta_p\beta_{rad}) \\
 \approx\epsilon^\prime(R_\tau)\Gamma_p(R_\tau)
    \frac{1}{2}\left(\frac{1}{\Gamma_p}
                 +\frac{\Gamma_p}{\Gamma_{rad}^2}\right).
\nonumber
\end{eqnarray}
The Compton temperature changes with radius as
\begin{equation}
 \frac{T_C(r)}{T_C(R_\tau)}=\frac{\epsilon^\prime(r)}{\epsilon^\prime(R_\tau)}
  \approx\frac{1}{2}\frac{\Gamma_p(R_\tau)}{\Gamma_p(r)}
         \left[1+\frac{\Gamma_p^2(r)}{\Gamma_{rad}^2(r)}\right],
\label{eq:TC}
\end{equation}
where $T_C(R_\tau)=T_r(R_\tau)$. Equation~(\ref{eq:TC}) completes the 
description of Compton energy exchange at the transparent stage.


\subsection{Numerical models}
\label{sec:pres}

\begin{figure}
\psfig{file=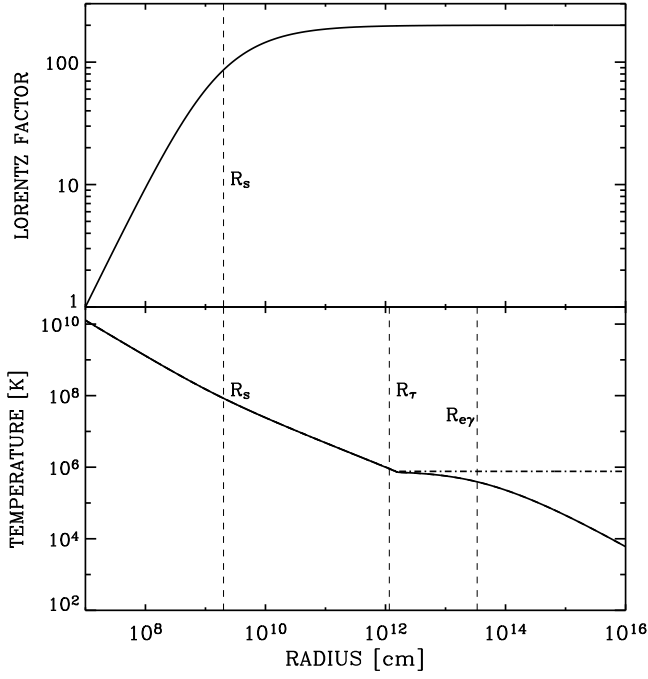,width=0.48\textwidth}
\caption[Neutron free radial evolution: $\eta=200$]
{ Lorentz factor (upper panel) and temperature (lower panel) as
functions of radius for a neutron-free outflow with $L=10^{52}$~erg/s, 
$\eta=200$ and $R_0=10^{7}$~cm. Electrons never decouple thermally from 
protons and their common temperature is shown by the solid curve.
Dashed-dotted curve shows the radiation temperature.}
\label{fig:p200}
\end{figure}
\begin{figure}
\psfig{file=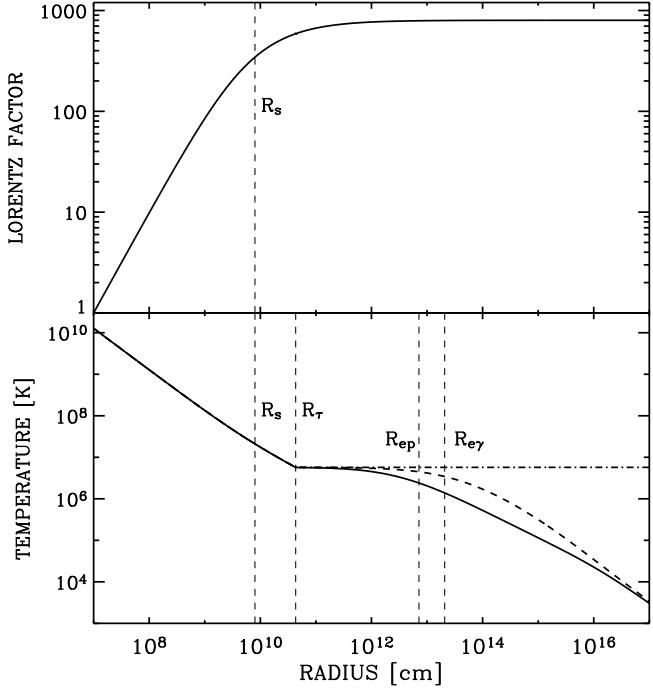,width=0.48\textwidth}
\caption[Neutron free radial evolution: $\eta=800$]
{{ Same as Fig.~\ref{fig:p200} but for $\eta=800$. 
$T_p$ and $T_e$ are shown by solid and dashed curves, respectively.
The thermal decoupling $T_e>T_p$ takes place at radii
$r\sim 10^{12}-10^{16}$~cm.}
\label{fig:p800}}
\end{figure}
\begin{figure}
\psfig{file=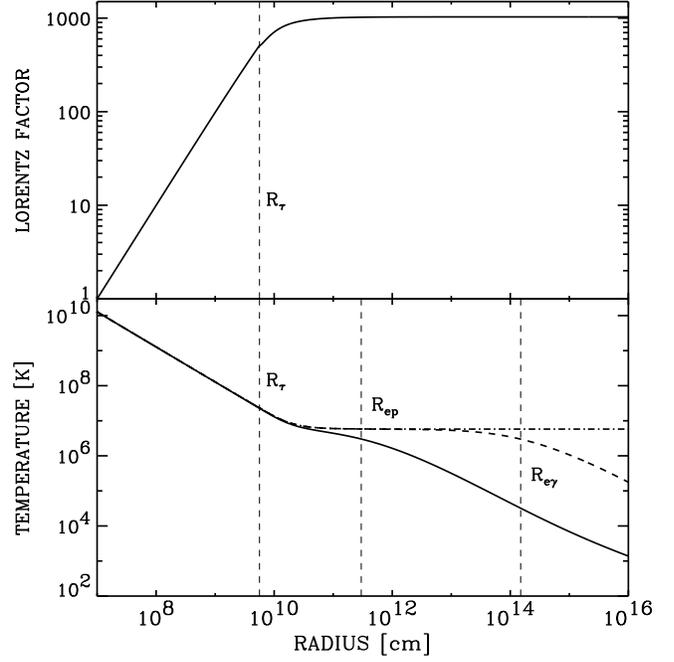,width=0.48\textwidth}
\caption[Neutron free radial evolution: $\eta=1800$]
{{ Same as Fig.~\ref{fig:p800} but for $\eta=8\times 10^{3}$. 
The outflow becomes transparent before the acceleration stage ends.}
\label{fig:p8e3}}
\end{figure}

Numerical models of outflows with $\eta=200$ and 800
are shown in Figures~\ref{fig:p200} and~\ref{fig:p800}. 
Both models have $\eta <\eta_{rad}$ and
the outflow Lorentz factor saturates before transparency.

Upper panels in Figures~\ref{fig:p200} and~\ref{fig:p800}  show the
evolution of the Lorentz factor.  As long as enthalpy exceeds
rest-mass energy ($x>1$), the outflow accelerates with $\Gamma_p
\propto r$.  When most of the internal energy is converted to kinetic
energy ($x<1$) $\Gamma_p$ tends to a constant asymptotic value
$\Gamma_{pf}=\eta$. The characteristic saturation radius $R_s$ is
where $x=1$; $\Gamma_p=\eta/2$ at this radius and
\be
  R_s=2^{-1/4}\eta R_0.
\label{eq:Rs}
\ee

Solutions for temperatures $T_r$, $T_e$ and $T_p$ are shown 
in the lower panels of Figures~\ref{fig:p200} and \ref{fig:p800}.
At very small radii where $kT\gsim m_ec^2=511$~keV, the number density 
of $e^\pm$ pairs is comparable to that of photons, the energy density is 
$(11/4)aT_r^4$ and the temperature is reduced by the factor of 
$(4/11)^{1/4}\approx 0.78$; this correction is neglected in the figures.
As long as the outflow is optically thick, 
all three components maintain a common temperature $T$ via Coulomb 
collisions and Compton scattering, and $T$ decreases adiabatically with index
$\hat{\gamma}=\frac{4}{3}$.
The outflow becomes transparent at radius
\be
R_\tau= \frac{L \sigma_T\,} {4\pi m_p c^3 \Gamma_{pf}^3}, 
  \qquad \eta<\eta_{rad}.
\label{eq:rt}
\ee
At the transparent stage the plasma is still tracking the temperature of 
(freely streaming) photons: $T_e\approx T_p\approx T_r\approx const$
until the electrons decouple either from radiation ($e$-$\gamma$ decoupling) 
or from the protons ($e$-$p$ decoupling).

The $e$-$\gamma$ decoupling occurs at a radius $R_{e\gamma}$ where the Compton 
time-scale (eq.~\ref{eq:tauc}) exceeds the expansion timescale 
$\tau_{exp}=R/\Gamma_p\,c$,
\be
R_{e\gamma}=\frac{2}{3\pi}\frac{L_\gamma\sigma_T}{m_ec^3\Gamma_{pf}^3}
           \approx\frac{2\sigma_T L}{3\pi\,m_ec^3\,\eta^{7/3}}\,
  \left(\frac{R_0}{R_{\tau}}\right)^{2/3}.
\label{eq:ref}
\ee
The $e$-$p$ decoupling may happen or not depending mostly on the value of 
$\eta$. The two possible cases are illustrated in Figures~~\ref{fig:p200} 
and ~\ref{fig:p800}:  

\noindent
(i) Electrons and protons are still coupled at $R_{e\gamma}$ 
($\tau_{ep}<\tau_{exp}$) and begin a common adiabatic cooling, 
$T_p=T_e\propto n_e^{2/3}$. They will not decouple later because 
$\tau_{ep}=const$ while $\tau_{exp}$ keeps increasing. This regime takes 
place at $\eta\lsim 650$ and is illustrated in Figure~\ref{fig:p200}.

\noindent
(ii) Electrons decouple from protons 
before $R_{e\gamma}$. The radius of $e$-$p$ decoupling is found from 
condition $\tau_{ep}=\tau_{exp}$. Using 
$T_e\approx T_r(R_\tau)=T_r(R_s)(R_\tau/R_s)^{-3/2}$ in 
equation~(\ref{eq:tauep}) for $\tau_{ep}$ (and neglecting the small term 
$kT_p/m_pc^2$), we get
\be
R_{ep}\simeq\left(68\pi\,m_pc^4\right)^{-1}
\frac{L}{\eta^{3}\,T_s^{3/2}}\left(\frac{ R_{\tau}} {R_s}\right),
\ee
where $T_s=2T_0/\eta$ is the temperature at the saturation radius.
At $R_{ep}<r<R_{e\gamma}$, $T_e\approx T_r>T_p$. At $r>R_{e\gamma}$
the electrons cool down adiabatically and $\tau_{ep}\approx const$ while 
$\tau_{exp}$ keeps increasing. Therefore, electrons and protons eventually 
regain the thermal coupling.
This regime takes place at high $\eta\gsim 650$ and is illustrated in 
Figure~\ref{fig:p800}.

Figure~\ref{fig:p8e3} shows a model with $\eta=8\times 10^{3}>\eta_{rad}$. 
In this case, the outflow acceleration continues after the transparency 
radius
\be
R_{\tau}\simeq \,R_0\, \eta_{rad}\,\left(\frac{\eta_{rad}}{\eta}\right)^{1/3},
 \qquad \eta>\eta_{rad},
\label{eq:rtauac}
\ee
and $\Gamma_p$ saturates at 
\be
\Gamma_{pf}\approx\eta_{rad}\,
\left(\frac{\eta}{\eta_{rad}}\right)^{1/9}<\eta,
\label{eq:gpftr}
\ee
after a few $R_\tau$. Since density is lower than in the previous
examples, the $e$-$p$ decoupling happens much earlier than the 
$e$-$\gamma$ decoupling ($R_{ep}\ll R_{\gamma}$).

In summary, the thermal evolution of a neutron-free outflow is dominated by 
adiabatic cooling. The opaque outflow has a common temperature
$T_p=T_e=T_r$ which decreases as $r^{-1}$ during the acceleration 
stage and as $r^{-2/3}$ during the coasting stage.
Then, after an intermediate stage where the details of coupling between 
$e$, $p$, and radiation are important, the outflow is again described
by a simple adiabatic law $T_p\approx T_e\propto r^{-4/3}$. 
The presence of neutrons will change this picture.


\section{Neutron-loaded outflow}
\label{sec:npfire}

We now consider an outflow with a neutron component and denote 
its initial neutron richness (neutron-to-proton ratio) by $\xi_0$.
All other assumptions are the same as in section~\ref{sec:pfire}. 
In particular, we consider spherical expansion driven 
by radiation pressure. 
Neutrons and protons are injected at $r=R_0$ with an initial density 
ratio $\xi=\xi_0$, and then $\xi$ evolves with radius because neutrons 
continuously $\beta$-decay into protons, 
$$
 n \rightarrow p+e+\bar{\nu}. 
$$ 
The mean life-time of neutrons in their rest frame is 
$\tau_{\beta}\approx 900$~s, and the corresponding mean radius of 
$\beta$-decay is 
\be
R_{\beta}=\int_{0}^{\tau_{\beta}}{c\beta_n\Gamma_n\,d\tau}\simeq 
\,8\times 10^{15}\left(\frac{\Gamma_{nf}}{300}\right) {\rm ~cm},
\label{betar}
\ee
where $\Gamma_n$ is the neutron Lorentz factor and $\beta_n\approx 1$;
$\Gamma_{nf}$ is the final value of $\Gamma_n$ achieved at 
$r\sim (10^2-10^3)R_0\ll R_\beta$.
The neutron population is gradually depleted as $\exp(-r/R_\beta)$
and the $n/p$ ratio (measured in the fixed lab frame) evolves with 
radius as 
\begin{equation}
  \xi\equiv\frac{\Gamma_n n_n}{\Gamma_pn_p}
  =\frac{\xi_0e^{-r/R_{\beta}}}{1+\xi_0\left(1-e^{-r/R_{\beta}}\right)},
\label{eq:xin}
\end{equation}
where $n_n$ and $n_p$ are proper densities of the neutron and proton 
components.

\subsection{Outflow acceleration and $n$-$p$ coupling}
\label{sec:npdec}

The total luminosity of the outflow (cf. eq.~\ref{eq:ce}) now includes 
the contribution from neutrons,
\be
L= 4\pi r^{2}c 
\left[\beta_p \Gamma_p^{2} \left(\frac{4}{3} a\,T_r^{4}+\rho_p\,c^2
   \right)+\beta_n \Gamma_n^{2}\rho_nc^2 \right]=const,
\label{eq:cen}
\ee
where $\rho_p=n_pm_p$ and $\rho_n=n_nm_n$;
the thermal energy of the plasma and neutrons has been neglected 
compared to their rest-mass energy.
We allow here the neutron component to have a different Lorentz factor 
$\Gamma_n$, which will be close to $\Gamma_p$ as long as the $n$-$p$ 
coupling is efficient. 

The baryon outflow rate is given by
\be
\dot{M}=4\,\pi \,r^{2}\,c\left(\beta_p \Gamma_p \rho_p
                          +\beta_n \Gamma_n \rho_n\right)=const.
\label{eq:cmn}
\ee

The outflow expansion is accompanied by adiabatic cooling which 
determines $T_r(r)$.
Expansion of volume can be described by the decrease of number
density of {\it original} protons injected at the base of the outflow.
We denote this density by $n$ and distinguish it from the total proton
density $n_p$ that includes the decayed neutrons. They are related by
\be
  n_p=n\,\frac{1+\xi_0}{1+\xi}.
\label{eq:npn}
\ee
The radiation temperature at the optically thick stage obeys 
equation~(\ref{eq:adlaw}). We neglect destruction of neutrons 
at this stage (see section~\ref{sec:inelastic}) and assume in this section 
$\xi=\xi_0$ and $n=n_p$. The difference between $n_p$ and $n$ will become
significant at larger radii comparable to $R_\beta$.

Acceleration of the optically thick outflow is described by 
equations~(\ref{eq:cen}), (\ref{eq:cmn}) and (\ref{eq:adlaw}), 
from which we derive
\be
  \frac{d\Gamma_p}{dr}=\frac{\Gamma_p}{r}\frac{2x}{2x+3}
        -\frac{d\Gamma_n}{dr}\frac{3\xi_0}{2x+3},
\label{eq:dgp}
\ee
where 
\be
\nonumber
  x=\frac{4}{3}\frac{aT_r^4}{n_pm_pc^2}=x_0\left(\frac{n}{n_0}\right)^{1/3},
   \qquad x_0=\frac{\eta(1+\xi_0)}{\Gamma_0}-1, 
\ee
\be
  x(r)=x_0\left(\frac{r}{R_0}\right)^{-2/3}
             \left(\Gamma_p\beta_p\right)^{-1/3}.
\label{eq:x1}
\ee
The first term on the right-hand side of equation~(\ref{eq:dgp}) describes 
the acceleration by radiation pressure; the second term describes 
the deceleration caused by transfer of momentum to neutrons. 

As long as the collisional $n$-$p$ coupling is strong, 
$\Gamma_n\approx\Gamma_p$, equation~(\ref{eq:dgp}) yields
\be
 \frac{d\Gamma_p}{dr}=\frac{\Gamma_p}{r}\frac{2x}{2x+3(1+\xi_0)}.
\label{eq:dgp1}
\ee
Acceleration of coupled $n$ and $p$ begins to saturate when 
$x\approx 1+\xi_0$ at a radius 
\be
 R_{sb}\approx\eta R_0, \qquad \Gamma_p(R_{sb})\approx \frac{\eta}{2}.
\label{eq:Rsb}
\ee
At that point $\rho_bc^2\approx aT_r^4$ where $\rho_b \approx \rho_p+\rho_n$. 
In the case of $\xi_0\gg 1$, acceleration saturates when $x$ is still
large because the protons are coupled to a large 
number of neutrons and have a lot of effective inertia. Decoupling 
from neutrons at $r\lsim R_{sb}$ would allow the protons to accelerate 
further and reach Lorentz factors $\Gamma_p\sim\eta(1+\xi_0)$ if the
outflow remains optically thick (Fuller at al. 2000). 

Description of $n$-$p$ decoupling will require one more equation that 
specifies momentum exchange between neutron and proton components.
Neutrons accelerate because they 
collide with the accelerating protons.
In its rest frame, a neutron experiences $\Gamma_{rel}n_p\sigma_{np}c$
collisions per second\footnote{To the first approximation the rate of 
$n$-$p$ collisions $<\sigma v>$ does not depend on the relative velocity 
$v$ of colliding particles and one may take $<\sigma v>=<\sigma_{np}c>$.}
where 
$\sigma_{np}\approx 3\times 10^{-26}$cm$^{2}$ and
\be
  \Gamma_{rel}=\Gamma_p\Gamma_n\left(1-\beta_p\beta_n\right)
              \approx \frac{1}{2}\left(\frac{\Gamma_p}{\Gamma_n}+
                                       \frac{\Gamma_n}{\Gamma_p}\right)
\label{eq:Grel}
\ee
is the Lorentz factor of the neutron component relative to the plasma
component.
Assuming isotropic scattering in the center-of-momentum frame, the mean 
momentum gained by the neutron per collision equals 
$\tilde{p}_p=\mu v_{rel}\Gamma_{rel}$ where 
$\mu=m_pm_n/(m_n+m_p)\approx m_p/2$ is the reduced mass. 
The momentum gained by a neutron during time $d\tilde{t}$ is
\be
  d\tilde{p}_n=\frac{1}{2}\,n_p\Gamma_{rel}^{2}\sigma_{np}\beta_{rel}
               m_pc^2\,d\tilde{t}.
\label{eq:dpdtn}
\ee
\no
The corresponding change of the neutron Lorentz factor in the lab frame is
found from the Lorentz transformation of 4-momentum
$d{\tilde p}_n^\alpha=(0,d\tilde{\mathbf p}_n)$,
\be
  d\Gamma_n=\Gamma_n\beta_n \frac{d\tilde{p}_n}{m_n\,c},
\ee
\no
and one gets
\be
\frac{d\Gamma_n}{dr}=\frac{1}{2}\,n_p\Gamma_{rel}^{2}\beta_{rel}\sigma_{np}.
\label{eq:dgn}
\ee
Equation~(\ref{eq:dgn}) closes the set of dynamic equations describing 
the outflow acceleration at the optically thick stage. 

The collisional coupling between the proton and neutron components is
friction, which inevitably dissipates energy and heats both components. 
This heating will be included in the plasma thermal balance below
(section 3.5).

\subsection{$n$-$p$ decoupling and transparency}

As long as the $n$-$p$ coupling is efficient, $\Gamma_p\approx \Gamma_n$,
the relative velocity $\brel\ll 1$ may be evaluated using 
equations~(\ref{eq:dgp1}) and (\ref{eq:dgn}),
\begin{eqnarray}
\nonumber
 \brel & \approx &\frac{2\Gamma_p}{n_p\sigma_{np}r}\frac{2x}{2x+3(1+\xi_0)} \\
       & = &\frac{2\Gamma_p}{n_p\sigma_{np}r}\frac{1}{1+(9/8)\rho_bc^2/aT_r^4},
\label{eq:brel}
\end{eqnarray}
The decoupling of $\Gamma_n$ from $\Gamma_p$ may happen at $r\lsim R_{sb}$
where $aT_r^4>\rho_bc^2$, and according to equation~(\ref{eq:brel})
\be
 \brel\approx\frac{2\Gamma_p}{n_p\sigma_{np} r}, \qquad  r<R_{sb}. 
\label{eq:brel1}
\ee
The decoupling condition is expressed by setting $\brel=1$ in
equation~(\ref{eq:brel1}). The $\brel$ approaches unity at $r\lsim R_{sb}$
if the outflow has a sufficiently high $\eta$,
\be
\label{eq:etastar}
  \eta>\eta_*\approx\left[\frac{L\sigma_{np}}
                      {4\pi
R_0m_pc^3(1+\xi_0)}\right]^{1/4}=\frac{4.8\times 10^{2}}{(1+\xi_0)^{1/4}}
\left(\frac{L_{52}}{R_{0,7}}\right)^{1/4}.
\label{eq:eta*}
\ee
Then the decoupling radius is given by
\be
  R_{np}\approx\left[\frac{L\sigma_{np}}
                     {8\pi R_0\eta m_pc^3(1+\xi_0)}\right]^{1/3}\,R_0,
\label{eq:Rnp}
\ee
and the neutron Lorentz factor at decoupling is
\be
\label{eq:Gnf}
  \Gamma_{nf}\approx \frac{R_{np}}{R_0}\approx\frac{\eta_*^{4/3}}{\eta^{1/3}},
   \qquad \eta > \eta_*.
\ee
No significant decoupling $\Gamma_p>\Gamma_n$ happens for $\eta<\eta_*$. 
This is so despite the fact that $\eta<\eta_*$ does not exclude 
$\rho_pc^2<aT_r^4<\rho_b c^2$ at $r>R_{sb}$, so there may be enough 
energy in radiation to accelerate protons to $\Gamma_p>\Gamma_{nf}$. 
A detailed analysis and numerical models show that if $\brel$ is small 
at $r\lsim R_{sb}$, it remains small at $r>R_{sb}$. Therefore,
to a good approximation, $\eta>\eta_*$ may be taken as a true condition 
for decoupling of $\Gamma_p$ from $\Gamma_n$. Hereafter we focus on 
$\eta>\eta_*$.

The decoupled neutrons do not affect the remaining $e$-$p$-$\gamma$ outflow 
until a much larger radius where $\beta$-decay becomes important (see below).
Therefore saturation of proton acceleration and transition to 
transparency may be described as if there were no neutrons.
This `neutron-free' outflow has luminosity
\be
  \hat{L}=L-\Gamma_{nf}\dot{M}c^2\,\frac{\xi_0}{1+\xi_0},
\ee
and mass outflow rate
\be
  \hat{\dot{M}}=\frac{\dot{M}}{1+\xi_0}.
\ee
Equations of section~2 then apply if one replaces 
$L\rightarrow \hat{L}$, $\dot{M}\rightarrow\hat{\dot{M}}$ and 
$\eta\rightarrow\hat{\eta}=\hat{L}/\hat{\dot{M}}c^2$. 

The outflow is still opaque to radiation at the $n$-$p$ decoupling 
radius. This follows from the small ratio of cross sections 
$\sigma_{np}/\sigma_T\sim 1/20$. If $e^\pm$ cascade initiated by $\pi^0$ 
is neglected (see DKK99a and section 3.7 below), transparency comes soon
after $R_{np}$. In this case, one can show that an outflow with 
$\Gamma_p>\Gamma_n$ ($\eta>\eta_*$) becomes transparent before saturation 
of $\Gamma_p$. This may be seen from the following relation,
\be
  \frac{\hat{\eta}}{\hat{\eta}_{rad}}
  =\left[1+\xi_0\left(1-\frac{\Gamma_{nf}}{\eta}\right)\right]^{3/4}
   \left(\frac{\sigma_T}{\sigma_{np}}\right)^{1/4}\,\frac{\eta}{\eta_*},
\ee
where (cf. eq.~\ref{eq:etamax})
\be
 \hat{\eta}_{rad}=\left(\frac{\hat{L}\sigma_T}{4\pi R_0 m_pc^3}\right)^{1/4}.
\ee 
$\eta>\eta_*$ implies $\hat{\eta}>\hat{\eta}_{rad}$ and hence $\Gamma_p$ 
does not reach the maximum possible value $\Gamma_{p,\max}=\hat{\eta}$. 
The outflow becomes transparent and $\Gamma_p$ saturates at 
(cf. eq.~\ref{eq:gpftr})
\be
\Gamma_{pf}\approx\hat{\eta}_{rad}\,
\left(\frac{\hat{\eta}}{\hat{\eta}_{rad}}\right)^{1/9}.
\label{eq:gpftr1}
\ee
It is not much larger than $\Gamma_{nf}$. The $e^\pm$ cascade 
initiated by $\pi^0$ decay may prolong the opaque stage and increase
$\Gamma_{pf}$.


\subsection{Deceleration by $\beta$-decay}
\label{sec:beta}

Next, we consider larger radii $r>R_\tau$ when the outflow is composed
of decoupled radiation with luminosity $L_\gamma$, decoupled neutrons
and plasma.  The neutrons are no longer a fluid; they retain a 
mildly-relativistic velocity dispersion acquired at their last collisions 
at $r\sim R_{np}$, which implies that $\Gamma_n$ varies by a factor $< 2$.
We will neglect this dispersion and assume that all neutrons have equal
Lorentz factors $\Gamma_n$ after decoupling.

The neutrons continuously $\beta$-decay in the outflow and create a 
source of protons and electrons moving with respect to the plasma frame
with a Lorentz factor $\Gamma_{rel}$. This beam shares momentum with 
the plasma, heats it and reduces the plasma Lorentz factor $\Gamma_p$. 

The outflow luminosity at the transparent stage may be approximated as 
\be
  L=4\pi r^2c\left[\beta_p\Gamma_p^2(\rho_pc^2+h_p)
                  +\beta_n\Gamma_n^2\rho_n c^2\right]+L_\gamma=const.
\label{eq:L1}
\ee
We include here the enthalpy of proton component $h_p=U_p+P_p$ for 
completeness. As we shall see the plasma is heated to 
a high temperature at large radii $r\sim R_\beta$, however, $h_p$ 
remains smaller than $\rho_pc^2$. The enthalpy is related to proton 
temperature by 
\be
  h_p=\frac{\hat{\gamma}}{\hat{\gamma}-1}n_pkT_p,
\label{eq:hp}
\ee
where $\hat{\gamma}$ is the adiabatic index of the proton component,
which is between $4/3$ and $5/3$. From equation~(\ref{eq:cmn}),
(\ref{eq:L1}) and using the first law of thermodynamics (eqs.~\ref{eq:Ilaw}
and ~\ref{eq:UP}) we derive,
\begin{eqnarray}
\nonumber
  \frac{d\Gamma_p}{dr}
  & = & \frac{1}{1+x_p(2-\hat{\gamma})}\left[
  \frac{\xi(\Gamma_n-\Gamma_p)}{R_\beta}
  +\frac{2(\hat{\gamma}-1)}{r}\Gamma_px_p \right. \\
\nonumber
  & - & \left. \frac{\Gamma_p\hat{\gamma}}{\rho_pc^2}\frac{dq_p}{dr}
  -\frac{\Gamma_px_p}{\hat{\gamma}}\frac{d\hat{\gamma}}{dr}\right] \\
  & + & \frac{\sigma_TL_\gamma}{16\pi r^2m_pc^3\Gamma_p^2} 
   \left(1-\frac{\Gamma_p^4}{\Gamma_{rad}^4}\right),
\label{eq:G1}
\end{eqnarray}
where $x_p=h_p/\rho_p c^2$.
The last term describes the radiative acceleration 
(see section 2.2.1).

\subsection{Summary of the dynamical model}

Our dynamical model of the neutron-loaded outflow may be summarized as 
follows. It is described by different equations before and after 
transparency, and the equations match at the transparency radius.
Before transparency, $r<R_\tau$, we neglect the decay of neutrons.
The evolution of $\Gamma_p$ and $\Gamma_n$ is found from 
equations~(\ref{eq:dgp}) and (\ref{eq:dgn}).

After transparency, $r>R_\tau$, we neglect the $n$-$p$ collisions and
assume $\Gamma_n=const$. We take into account the $\beta$-decay which 
becomes increasingly important at larger radii. The evolution of $\Gamma_p$ 
is described by equation~(\ref{eq:G1}). It includes the heating term 
$dq_p/dr$ which may be non-negligible at $r\sim R_\beta$. This term couples
the dynamics with the proton thermal balance which is discussed in 
section~3.5 below.

The two descriptions match at the beginning of the coasting phase
$\frac{d\Gamma_p}{dr}=0$ where neither $n$-$p$ collisions nor $\beta$-decay 
affects the outflow.

\subsection{Thermal balance}

Thermal evolution of electrons and protons obeys the first law 
of thermodynamics (eq.~\ref{eq:Ilaw}). Using equations~(\ref{eq:UP}) 
and (\ref{eq:npn}) we derive 
\begin{equation}
\left(1+\xi\right)\frac{d}{dr}
    \left[\frac{T_i}{(\hat{\gamma}-1)(1+\xi)}\right]
    =\frac{T_i}{n}\,\frac{dn}{dr}
     +\frac{1}{kn_p}\,\frac{dQ_i}{dt^\prime}\,\frac{1}{c\,\Gamma_p\,\beta_p}.
\label{eq:dten}
\end{equation}
Here $dQ_i/dt^\prime$ ($i=e,p$) are the heating rates of electrons and
protons.

The thermal balance of protons is significantly changed compared to the 
neutron-free case because of two effects: (i) frictional heating due 
to $n$-$p$ collisions and (ii) heating due to $\beta$-decay.

The rate of $n$-$p$ collisions per unit volume in the plasma frame is 
$$
 \dot{n}_{ep}=n_n^\prime n_p\sigma_{np}c,
$$
where $n_n^\prime=\Grel n_n$ is neutron density measured in the plasma frame.
The mean energy dissipated per collision is 
$(1/2)(\Gamma_{rel}-1)m_pc^2$ assuming that the relative bulk velocity is 
isotropized in collisions. This gives the frictional heating rate,
\be
  \frac{dQ_{np}}{dt^\prime}=\frac{1}{2}\,\Grel(\Grel-1) m_pc^3
                            \sigma_{np}n_n n_p.
\label{eq:qnp}
\ee

Heating also results from the gradual $\beta$-decay of the 
neutron component. The decay rate per unit volume is Lorentz invariant
and in the plasma frame (where the decay time is $\Gamma_{rel}\tau_\beta$) 
may be written as 
\be
  \dot{n}_\beta=\frac{n_n^\prime}{\Gamma_{rel}\tau_\beta}
               =\frac{n_n}{\tau_\beta}.
\ee
The decay products form an $e$-$p$ beam with velocity $v_{rel}$ in 
the plasma frame, which immediately dissipates its relative kinetic 
energy into heat. We will assume that this heat goes entirely to 
the proton component. The dissipated energy per decayed neutron is 
$(\Gamma_{rel}-1)m_pc^2$, which gives the heating rate,
\be
  \frac{dQ_\beta}{dt^\prime}=\left(\Gamma_{rel}-1\right)
                             m_pc^2\frac{n_n}{\tau_\beta}.
\label{eq:qb}
\ee

Protons also exchange energy with electrons via Coulomb collisions with 
rate  $dQ_{ep}/dt^\prime$ (eq.~22).
The net thermal balance of the proton component is then given by
\begin{eqnarray}
\nonumber
\frac{dq_p}{dt^\prime} &=&
  -\frac{dQ_{ep}}{dt^\prime} \\
 & + & \Grel n_n\left(\Gamma_{rel}-1\right)m_pc^2
   \left(\frac{1}{2}\,n_p\sigma_{np}c
        +\frac{1}{\Gamma_{rel}\tau_{\beta}}\right).
\label{eq:dqpn}
\end{eqnarray}
The $\beta$-decay weakly affects the plasma thermal balance as long as
$(\Gamma_{rel}\tau_\beta)^{-1}\ll n_p\sigma_{np}c$.
The decay becomes important at large radii, after the outflow becomes
transparent.

The thermal-balance equation for electrons is similar to the neutron-free
case (section~2),
\begin{equation}
 \frac{dQ_e}{dt^\prime}=\frac{3}{2}\,n_pk
              \frac{\left(T_C-T_e\right)} {\tau_c}
              +\frac{dQ_{ep}}{dt^\prime}.
\label{eq:dqen}
\end{equation}


\subsection{Destruction of neutrons by inelastic collisions}
\label{sec:inelastic}

We assume in this paper that neutron richness $\xi$ changes only
as a result of $\beta$-decay and neglect other channels of neutron
conversion to protons. In fact, near the decoupling radius $R_{np}$ 
the collisions between neutrons and protons become sufficiently 
energic to produce pions (DKK99a). Thus, some 
collisions are inelastic, which may destroy the neutron component.
 
Inelastic $n$-$p$ collisions may convert neutron 
to proton $n+p\rightarrow p+p+\pi^-$ as well as proton to neutron 
$n+p\rightarrow n+n+\pi^+$. These reactions have equal cross section
which has been measured down to the 140~MeV threshold 
(e.g. Daum et al. 2002). The reactions have equal rates and 
do not change neutron richness $\xi$.

Inelastic $n$-$n$ collisions may destroy the neutron component 
via reactions $n+n\rightarrow d+\pi^-$ and $n+n\rightarrow n+p+\pi^-$. 
These reactions have same cross sections as reactions 
$p+p\rightarrow d+\pi^+$ and $p+p\rightarrow n+p+\pi^+$, respectively, 
which have been studied in experiments (Shimizu et al. 1982). 
The cross section of $d\pi^-$ channel does not exceed 0.1 of the 
elastic cross section and is less important. The main reaction
of neutron destruction is $n+n\rightarrow n+p+\pi^-$. Its cross 
section becomes comparable to the elastic cross section when neutrons
collide with relative energy $E_{nn}>700$~MeV and quickly decreases
at smaller $E_{nn}$.

The rate of elastic $n$-$n$ collisions is $\xi$ times higher than the 
rate of $n$-$p$ collisions, which equals the expansion rate at decoupling. 
Therefore, the timescale of $n$-$n$ collisions at $r\sim R_{np}$ is 
$\xi^{-1}$ times shorter than the expansion time $\tau_{exp}$, and at 
large $\xi$ neutrons may be treated as Maxwellian gas. This gas is heated 
by $n$-$p$ collisions which dissipate energy 
$(\Grel-1)m_pc^2/2\approx m_pv_{rel}^{2}/4$ per collision, and the 
heating rate of neutrons is equal to that of protons (eq.~\ref{eq:qnp}). 
The neutron temperature $kT_n=(2/3)\bar{E}_n$ may be estimated by 
integrating the heating rate per neutron 
$d\bar{E}_n/dr\approx (m_pc^2/2r)\brel\approx(m_pc^2/2r)(r/R_{np})^3$ from 
$r=0$ to $r=R_{np}$. This gives a modest value $kT_n\sim 100$~MeV. 
A more accurate calculation, which includes adiabatic cooling of neutrons, 
may give even lower $T_n$. We conlcude that the mean relative energy of 
$n$-$n$ collisions, $E_{nn}=3kT_n$, is well below 700~MeV, so only a tail 
of the quasi-Maxwellian distribution will contribute to destruction of 
neutrons. 

Therefore most of $n$-$n$ collisions at $r\sim R_{np}$ are expected to
be elastic, and a small fraction $\zeta$ of these collisions will convert 
neutrons to protons.  
The lifetime of a neutron at $r\sim R_{np}$ is
$\tau_{n\rightarrow p}\approx\zeta^{-1}\xi^{-1}\tau_{exp}$.
An initially high neutron-to-proton ratio $\xi$ will be reduced by conversion
at decoupling if $\xi^{-1}\tau_{n\rightarrow p}<\tau_{exp}$, i.e. if the
initial $\xi$ exceeds $\zeta^{-1/2}$. 

The exact $\zeta$ and the corresponding upper bound $\xi_{max}$ may be 
found with detailed kinetic calculations of collisions.
Such calculations are not attempted here, and we consider an optimistic
range of $\xi<10$ which might remain intact after decoupling.
The case $\xi_0=4$ is chosen as a main example model.

\subsection{Pair cascade initiated by $\pi^0$}

Collisions between baryons near decoupling can produce 
neutral pions $\pi^0$ via reactions $p+p\rightarrow p+p+\pi^0$, 
$n+p\rightarrow d+\pi^0$ and $n+n\rightarrow n+n+\pi^0$.
Decay of $\pi^0$ initiates a pair cascade in the outflow (DKK99a).

The inelastic $p$-$p$ collisions are relatively rare at $\xi>1$,
especially when protons are cooled by Coulomb collisions with electrons.
Reaction $n+p\rightarrow d+\pi^0$ has a maximum cross section 
$\sim 0.1$ of the elastic cross section, and yields only $\sim 0.1$ $\pi^0$ 
per proton at decoupling.

The cross section of reaction $n+n\rightarrow n+n+\pi^0$ is $\sim 1/4$ 
times smaller than cross section of $n+n\rightarrow n+p+\pi^-$.
(These reactions have the same cross sections as the experimentally 
studied reactions $p+p\rightarrow p+p+\pi^0$ and $p+p\rightarrow p+n+\pi^+$,
respectively.) Hence, the timescale of $\pi^0$ production by a neutron 
is 4 times longer than $\tau_{n\rightarrow p}$ which is in turn longer 
than $\tau_{exp}$ (section~\ref{sec:inelastic}). This implies that 
up to $\sim 0.2$ $\pi^0$ is produced per neutron at decoupling, which 
translates to $\sim 1$ $\pi^0$ per electron.

A typical produced $\pi^0$ has energy of a few hundred MeV.
It immediately decays into two photons, and the high-energy photons are
absorbed by the thermal radiation
(DKK99a). As a result, a relativistic 
$e^\pm$ pair is created which upscatters thermal radiation, and the 
upscattered photons create a new generation of pairs. 
A maximum possible pair yield of the initiated cascade, $Y\approx 0.1$, 
would be achieved if the cascade were `saturated', i.e., if the upscattered
$\gamma$-rays always got absorbed by softer photons (Svensson 1987). 
Thus maximum 10 per cent of the $\pi^0$ energy could possibly be 
converted into pair rest mass, which corresponds to $\sim 30$ $e^\pm$
per injected $\pi^0$. The cascade is not, however, saturated: after a few 
generations, the upscattered $\gamma$-rays are able to escape the thermal 
radiation (DKK99a). Therefore, a more realistic pair yield
is a few per cent. It corresponds to roughly $10$ $e^\pm$ injected per
proton in the outflow.
Most of these pairs annihilate soon after injection at $r\sim R_{np}$.

The inelastic $n$-$n$ collisions also happen at $r>R_{np}$ with increasing
timescale $\propto n^{-1}\propto r^2$, and the pair creation remains 
significant until $r\sim 4R_{np}$. The created pairs increase the opacity
of the outflow and reduce the timescale of Coulumb interactions between
protons and electrons. In numerical models presented below this effect 
is neglected.


\subsection{Numerical models}
\label{sec:npres}

\begin{figure}
\psfig{file=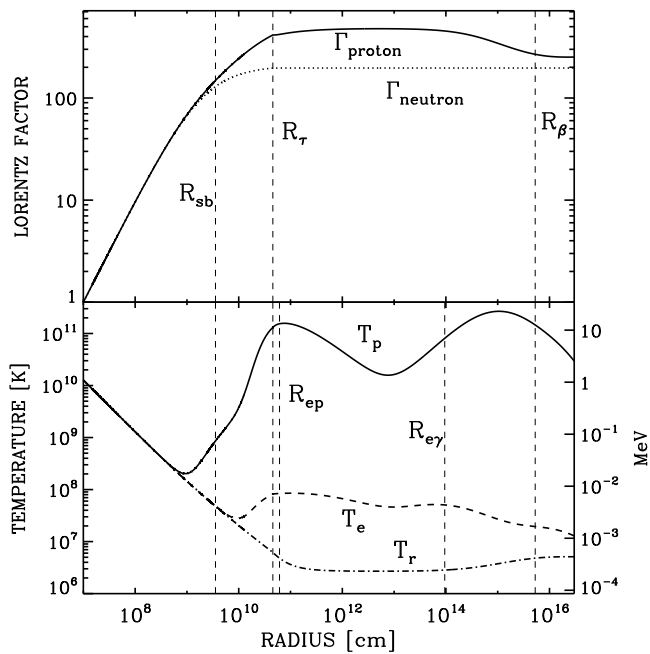,width=0.48\textwidth}
\caption[Radial evolution for $\xi=10$,$\eta=300$]
{Evolution of a neutron-loaded outflow with $\xi_0=4$, $\eta=300$, 
$L=10^{52}$~erg~s$^{-1}$, and $R_0=10^7$~cm.
{\it Upper panel}: Lorentz factors of proton and neutron components
(solid and dotted curves, respectively).
{\it Bottom panel}:  Temperatures of protons (solid curve), electrons 
(dashed curve) and radiation (dash-dotted curve). Vertical dashed lines
indicate characteristic radii $R_{sb}$, $R_\tau$, $R_{ep}$, $R_{e\gamma}$
and $R_\beta$ (see the text). The decoupling radius $R_{np}$ is close 
to $R_{sb}$.
\label{fig:g&tnp}}
\end{figure}

\begin{figure}
\psfig{file=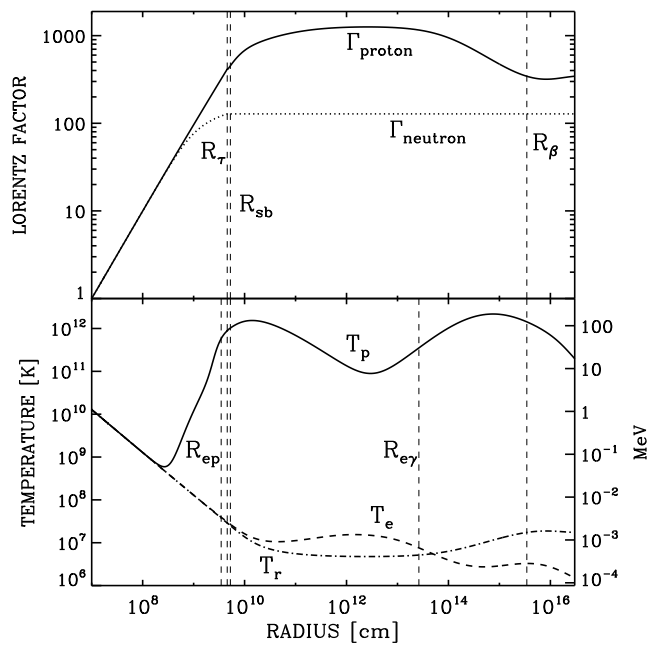,width=0.48\textwidth}
\caption[Radial evolution for $\xi=10$,$\eta=3e3$]
{Same as Fig.~\ref{fig:g&tnp} but for $\eta=3\times 10^{3}$. 
\label{fig:g&tnp3e3}}
\end{figure}

Figures~\ref{fig:g&tnp} and~\ref{fig:g&tnp3e3} show two numerical models
of neutron-loaded outflows with initial neutron-to-proton ratio $\xi_0=4$.
The first model has $\eta=300$ and the second model --- $\eta=3\times 10^3$.
In both cases, $\eta>\eta_*$ and the decoupling $\Gamma_p>\Gamma_n$ takes 
place. It is especially significant in the high-$\eta$ model.


The evolution of Lorentz factors $\Gamma_n$ and $\Gamma_p$ has three stages:
(i) Acceleration of both components which ends with different saturated 
values of $\Gamma_n$ and $\Gamma_p$. Note that neutrons decouple earlier
in the $\eta=3\times 10^3$ model and therefore have a smaller $\Gamma_n$ 
compared to the $\eta=300$ model. (ii) Coasting stage $\Gamma_n=const$ 
and $\Gamma_p=const$. (iii) Deceleration of the proton component by 
decaying neutrons. At about the same time the outflow begins to experience 
deceleration by an external medium. We do not study the external deceleration
in the present paper and stop our calculations at $r=3\times 10^{16}$~cm. 

The evolution of proton temperature $T_p$ has an initial decline followed 
by two pronounced peaks. The first peak happens at the end of neutron 
acceleration. It is caused by the strong friction between the $n$ and $p$ 
components near the decoupling radius $R_{np}$. The second peak 
accompanies the deceleration of protons at $r\lsim R_\beta$ ---
it is caused by absorption of decayed neutrons by the proton outflow. 
The overall thermal evolution is markedly different from the 
neutron-free case and we describe it below in more detail.

At the beginning of outflow expansion, the thermal evolution is similar 
to the neutron-free case: all components are thermally coupled at a common
temperature. The temperature is controlled by radiation (which strongly
dominates heat capacity of the outflow) and decreases according to 
adiabatic law with $\hat{\gamma}=4/3$. 

The behaviour changes when 
the Lorentz factor $\Gamma_p\approx\Gamma_n$ reaches a value about 50. 
Then the proton temperature decouples from $T_e$ and $T_r$ and begins to grow. 
This is the result of the frictional heating and the quickly increasing
relative velocity between $n$ and $p$ components. 
Only a fraction of the frictional heat is kept by baryons 
and a quasi-steady energy circulation is maintained in the 
accelerating outflow: radiation $\rightarrow$ relative kinetic energy
of the $n$ and $p$ components $\rightarrow$ baryonic heat $\rightarrow$ 
electrons $\rightarrow$ radiation.
The thermal balance of protons in this circulation controls
$T_p$, and $T_p$ quickly grows as the outflow approaches the $n$-$p$ 
decoupling radius.

After the decoupling radius, the $n$-$p$ collisions become rare and the 
frictional heating extinguishes. On the other hand, the protons get 
thermally decoupled from electrons because of a long timescale of Coulomb 
energy exchange.
The subsequent decrease of $T_p$ is mainly caused by adiabatic cooling.  

Adiabatic cooling continues until the heating by $\beta$-decay interferes
the thermal evolution. The timescale of adiabatic cooling in the proton frame 
is $r/c\Gamma_p$ which is proportional to $r$ during the coasting stage. 
By contrast, the timescale of $\beta$-decay is constant. Therefore the
heating of protons by decayed neutrons wins the adiabatic cooling at some 
point, and $T_p$ begins to grow as $T_p\propto r$. This growth continues 
until the outflow approaches $R_\beta$ where most of the neutrons decay.
The exponential extinction of neutrons implies that heating extinguishes 
at $r\sim R_\beta$. Then the outflow again cools adiabatically.

\begin{figure}
\psfig{file=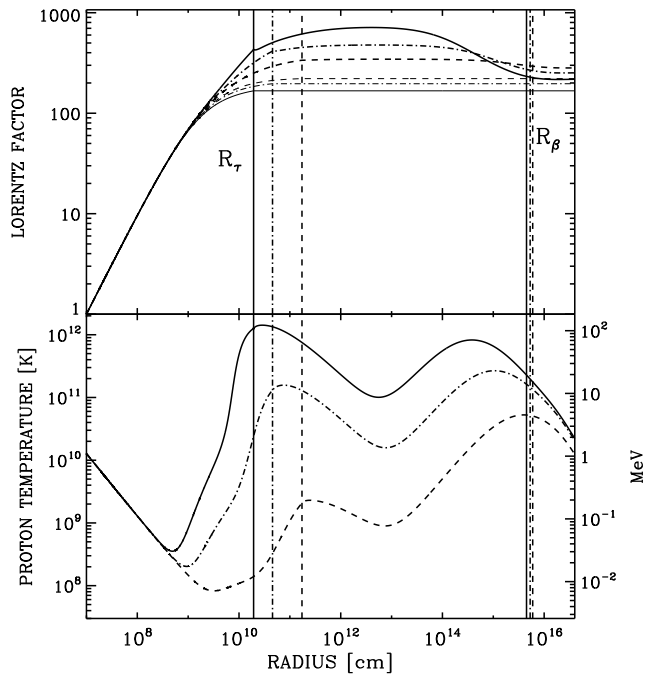,width=0.48\textwidth}
\caption[Radial evolution for different $\xi$]
{Evolution of outflows with the same parameters as in Fig.~4 but 
with different $\xi_0=1$ (dashed curves) and $\xi_0=10$ (solid curves). 
The model with $\xi_0=4$ from Fig.~4 is shown by the dash-dotted curves.
{\it Upper panel}: Lorentz factors of proton and neutron components.
{\it Lower panel}: proton temperature.
\label{fig:npxi}}
\end{figure}


\begin{figure}
\psfig{file=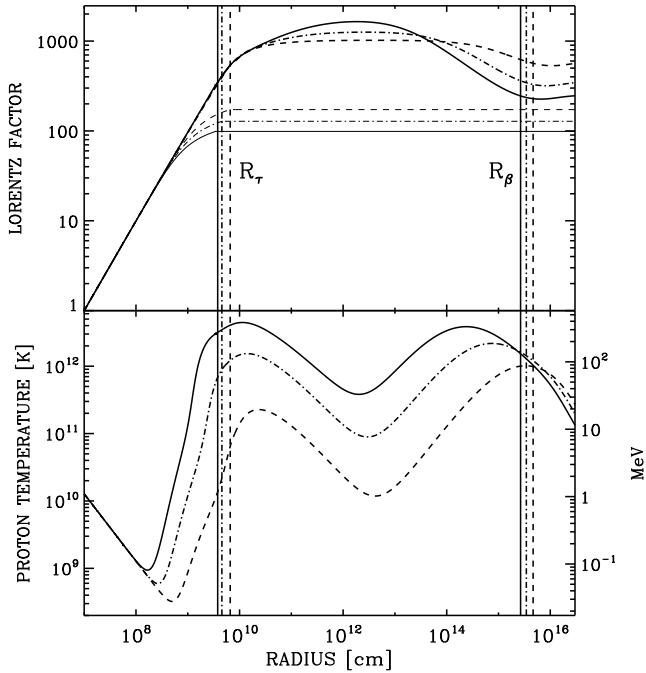,width=0.48\textwidth}
\caption[Radial evolution for different $\xi$ $\eta=3e3$] { Same as
Fig.~\ref{fig:npxi} but for outflows with $\eta=3 \times 10^3$. 
\label{fig:npxi2}}
\end{figure}

The two main effects of neutrons on the outflow --- heating and deceleration
of the proton component --- are especially strong at high $\xi_0$ and high 
$\eta$. This is illustrated in Figures~\ref{fig:npxi} and~\ref{fig:npxi2}
which show models with $\xi_0=1,4,10$ and $\eta=300,3\times 10^3$.
At high $\xi_0$ and/or $\eta$ the proton density of the outflow is 
low and the $n$-$p$ decoupling occurs early, leading to a large 
relative Lorentz factor $\Grel$ between the $n$ and $p$ components. 
The large kinetic energy of the relative motion is then available for 
dissipation.  

\begin{figure}
\psfig{file=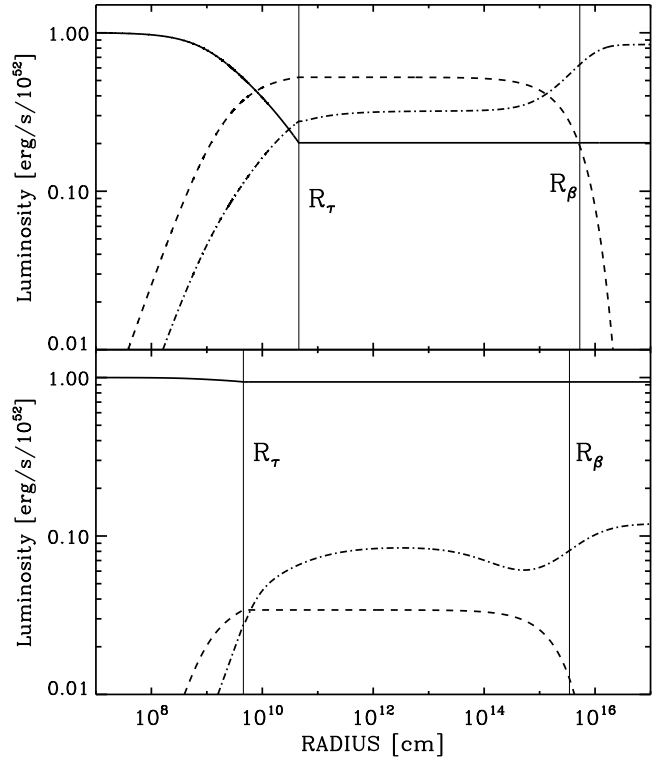,width=0.48\textwidth}
\caption[Luminosity evolution]
{Contributions to the energy outflow rate from radiation (solid curve), 
protons (dash-dotted curve) and neutrons (dashed curve). The sum of the
three luminosities $L_\gamma+L_p+L_n$ is constant and equal to 
$L=10^{52}$~erg~$^{s-1}$.
The upper panel corresponds to the model shown in Fig.~\ref{fig:g&tnp}, 
and the lower panel --- to the model shown in Fig.~\ref{fig:g&tnp3e3}.
\label{fig:lk}}
\end{figure}

We also note that outflows with very high $\eta$ and $\xi_0$ quickly 
become transparent and most of their energy is carried away by thermal 
radiation. This is illustrated by Figure~\ref{fig:lk} which shows three 
components of the outflow luminosity: radiation, protons, and neutrons.
The radiation luminosity is given by
$L_{\gamma}= 4\pi \,r^{2} c\beta_p \Gamma_p^{2} \left(\frac{4}{3}
a\,T_r^{4}\right)$ for $r<R_{\tau}$ and remains constant after $R_\tau$.
The kinetic luminosities of protons and neutrons are 
$L_p=4\pi \,r^{2}c\beta_p \Gamma_p^{2}\rho_p\,c^2$ and 
$L_n=4\pi \,r^{2}c\beta_n \Gamma_n^{2}\rho_n\,c^2$.
The sum of three contributions remains constant and
equals the total luminosity $L=10^{52}$~erg/s. 
One can see that the outflow with 
$\xi_0=4$ 
and $\eta=300$ emits about
30 per cent of its energy at the transparency radius, and the outflow
with $\eta=3\times 10^3$ --- more than 90 per cent.

All numerical models shown in Figures~4-8 assumed that the outflow 
starts to accelerate at radius $R_0=10^7$~cm.
This is a reasonable assumption if the outflow forms near a stellar-mass 
compact object and expands either spherically or conically with a 
constant opening angle $\theta_j>1/\Gamma_p$.
It is likely, however, that the free conical expansion is preceeded
by a collimation stage (as envisioned by, e.g., collapsar model of
MacFadyen \& Woosley 1999). During the collimation stage 
$\theta_j(r)$ decreases and the jet is hardly accelerated inside the
progenitor star until it reaches the break-out radius $R_0$ and starts 
free expansion with $\theta_j(r)=const$ and acceleration $\Gamma_p\propto r$.
Then our spherical/conical model applies with a large effective $R_0$.

To see how the outflow dynamics is changed by collimation
we have calculated numerical models with $R_0=10^7,10^8,10^9$~cm.
The results are shown in Figure~9.
When $R_0$ is large, a given Lorentz factor is reached at a smaller 
density of the outflow. As a result, neutrons decouple at a smaller 
Lorentz factor $\Gamma_n$ and the outflow develops a larger $\Grel$,
leading to strong heating.
This trend is observed in Figure~9. Outflows with large $R_0$ also 
have smaller $\Gamma_p$ and most of their energy is carried away 
by thermal radiation released at the transparency radius.

We also note that explosions with large $R_0$ have larger decoupling
radii $R_{np}$ and $R_\tau$, and a smaller decay radius $R_\beta$.
Therefore the two peaks of frictional and $\beta$-decay heating are 
closer to each other, and the stage of adiabatic cooling between them 
shortens. At even larger $R_0\gsim 10^{10}$ the two peaks would overlap.


\begin{figure}
\psfig{file=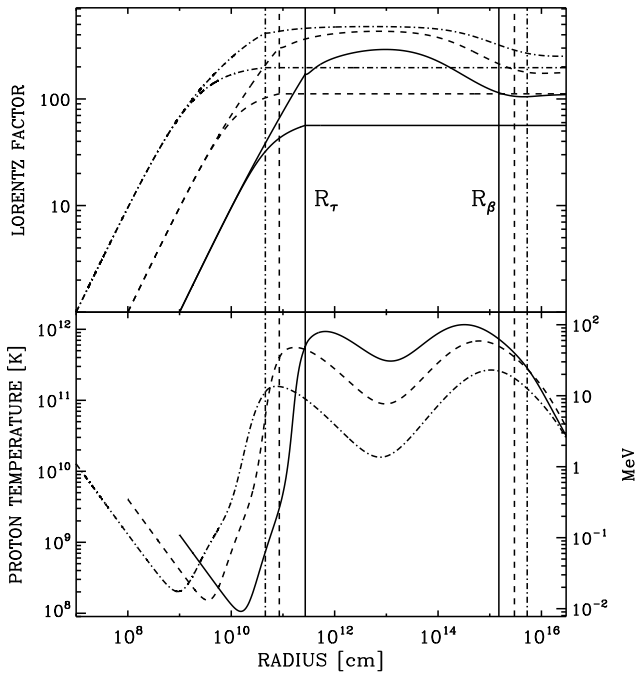,width=0.48\textwidth}
\caption[Radial evolution for different $\eta$] 
{ Evolution of outflows with the same parameters as in Fig.~4 but
with different $R_0=10^{8}$~cm (dashed curves) and $R_0=10^{9}$~cm 
(solid curves).
The model with $R_0=10^7$~cm from Fig.~4 is shown by the dash-dotted curves.
{\it Upper panel}: Lorentz factors of proton and neutron components.
{\it Lower panel}: proton temperature.
\label{fig:r0}}
\end{figure}

\section{Discussion}
\label{sec:discc4}

In this paper we developed the theory of relativistic neutron-loaded 
outflows. The presence a neutron component significantly
affects the early dynamics of GRB explosions. In particular, the plasma
temperature is increased by many orders of magnitude, and this heating 
can compete with other heating mechanisms such as internal shocks or
dissipation of magnetic fields in the outflow.

The effects of neutrons are pronounced in outflows with $\eta>\eta_*$
given by equation~(\ref{eq:etastar}), whose typical value is several 
hundred. Then neutrons and protons develop a substantial relative Lorentz 
factor which leads to strong heating and momentum exchange when neutrons 
decay. We also note that neutron-rich outflows with very high 
$\eta\gg\eta_*$ lose most of their energy at the photosphere: their thermal 
radiation carries most of the energy at the moment of transparency. 
Therefore, the neutron effects may be especially strong in GRBs with a 
significant thermal component in the radiation spectrum (a number of such 
GRBs have been identified recently, see e.g. Ghirlanda 2003; Ryde 2004).

The calculations in this paper focused on the simplest model of a uniform 
hydrodynamic outflow with weak magnetic fields. We did not consider, 
for instance, internal shocks (e.g. Rees \& M\'esz\'aros 1994) and 
possible pair creation by nonthermal $\gamma$-rays generated in the outflow. 
Pair creation may extend the optically thick stage of expansion, and the 
trapped radiation may convert its energy more efficiently into bulk kinetic 
energy of the plasma.
 A large-scale magnetic field may gradually collimate the outflow so 
that the conical geometry of expansion does not apply (Vlahakis, Peng
\& K\"onigl 2003).

The $\beta$-decay is likely to affect the development of internal shocks in 
the outflow. The shocks are caused by a non-uniform profile of the Lorentz 
factor, and the drag effect of decayed neutrons tends to smoothen this 
profile. The fastest portions of the outflow are more effectively 
decelerated and the initial contrast of Lorentz factors may be 
substantially reduced already at $r\sim 10^{14}$~cm. This effect constrains 
the dissipation efficiency of the shocks, which is sensitive to the contrast
of Lorentz factors (see Fig.~3 in Beloborodov 2000). In addition, the high 
temperature of the outflow heated by $\beta$-decay may prevent development 
of the shocks. We defer a detailed study of these effects to a future work. 

The impact of neutrons on the prompt burst and its afterglow provides a 
unique opportunity to link the observed emission with physical conditions
in the central engine of the explosion. The very presence of neutrons is 
a signature of an extremely hot and dense engine. Observable effects of 
neutrons may shed light on the mechanism of GRB trigger.


\section*{Acknowledgments}
AMB was supported by NASA grant NAG5-13382 and the Alfred P. Sloan Fellowship.
EMR and MB thank the KITP at Santa Barbara for hospitality during the 
completion of this work.

\end{document}